\documentclass[aps,prb,twocolumn,showpacs,preprintnumbers,superscriptaddress,amsmath,amssymb]{revtex4}

\usepackage{color}
\usepackage{graphicx}% Include figure files
\usepackage{dcolumn}% Align table columns on decimal point
\usepackage{bm}% bold math

\begin{document}

\title{c-axis magnetotransport in CeCoIn$_{5}$}

\author{A. Malinowski}
\affiliation{Materials Science and Technology Division, Los Alamos
National Laboratory, Los Alamos, New Mexico 87545, USA\\}
\affiliation{Institute of Physics, Polish Academy of Sciences, 02
668 Warsaw, Poland}
\author{M. F. Hundley}
\author{C. Capan}
\altaffiliation[Present address: ]{Department of Physics and
Astronomy, Louisiana a State University, Baton Rouge, LA 70803.}
\author{F. Ronning}
\author{R. Movshovich}
\author{N. O. Moreno}
\author{J. L. Sarrao}
\author{J. D. Thompson}
\affiliation{Materials Science and Technology Division, Los Alamos
National Laboratory, Los Alamos, New Mexico 87545, USA\\}

\date{\today}

\begin{abstract}
We present the results of out-of-plane electrical transport
measurements on the heavy fermion superconductor CeCoIn$_{5}$ at
temperatures from 40 mK to 400 K and in magnetic field up to 9 T.
For $T <$ 10 K transport measurements show that the zero-field
resistivity $\rho_{c}$ changes linearly with temperature and
extrapolates nearly to zero at 0 K, indicative of non-Fermi-liquid
(nFL) behavior associated with a quantum critical point (QCP). The
longitudinal magnetoresistance (LMR) of CeCoIn$_{5}$ for fields
applied parallel to the c-axis is negative and scales as
$B/(T+T^{*})$ between 50 and 100 K, revealing the presence of a
single-impurity Kondo energy scale $T^{*} \sim 2$ K. Beginning at 16
K a small positive LMR feature is evident for fields less than 3
tesla that grows in magnitude with decreasing temperature. For
higher fields the LMR is negative and increases in magnitude with
decreasing temperature. This sizable negative magnetoresistance
scales as $B{^2}/T$ from 2.6 K to roughly 8 K, and it arises from an
extrapolated residual resistivity that becomes negative and grows
quadratically with field in the nFL temperature regime. Applying a
magnetic field along the c-axis with B $>$ B$_{c2}$ restores
Fermi-liquid behavior in $\rho_{c}(T)$ at $T$ less than 130 mK.
Analysis of the $T{^2}$ resistivity coefficient's field-dependence
suggests that the QCP in CeCoIn$_{5}$ is located \emph{below} the
upper critical field, inside the superconducting phase. These data
indicate that while high-$T$ c-axis transport of CeCoIn$_{5}$
exhibits features typical for a heavy fermion system, low-$T$
transport is governed both by spin fluctuations associated with the
QCP and Kondo interactions that are influenced by the underlying
complex electronic structure intrinsic to the anisotropic
CeCoIn$_{5}$ crystal structure.

\end{abstract}

% PACS, the Physics
% and Astronomy Classification Scheme.
%%%%%%%%%%%%%%%%%%%%%%%%%%%%%%%%%%%%%%%%%%%%%%%%%%%%%%%%%%%%%%%%%
%%% 71.27.+a Strongly correlated electron systems; heavy fermions
%%% 73.43.Qt Magnetoresistance
%%% 75.40.-s Critical-point effects, specific heats, short-range
%%% order
%%% 74.70.Tx Heavy-fermion superconductors
\maketitle

\section{\label{Intro}Introduction}
Extensive evidence for departures from the temperature-dependencies
characteristic of Fermi-liquid (FL) behavior in the thermodynamic
properties of $d$- and $f$-metals has been collected over the last
fifteen years.\cite{Stewart01} A key observation is that both the
Sommerfeld coefficient, $\gamma=C/T$, and the Pauli susceptibility
$\chi$ increase with decreasing temperature and show no sign of
entering a $T$-independent FL regime down to the lowest achievable
temperatures. Another primary indication of non-Fermi-liquid (nFL)
behavior in heavy fermion (HF) systems is provided by the
low-temperature non-quadratic (in some systems close to linear)
temperature dependence exhibited by the resistivity in these
compounds.\cite{Maple95} This transport behavior may link the
heavy-fermion compounds with the copper oxide superconductors where
a $T$-linear resistivity is observed over a wide temperature range
that often extrapolates to zero at $T =
0$.\cite{Gurvitch87,Stormer88}

Theories trying to explain nFL behavior can be divided into three
categories:\cite{Stewart01} (1) models that describe the behavior
expected near a quantum critical point (QCP), (2) multichannel
single-impurity Kondo models, and (3) models based on disorder.
These general categories are not exclusive; an anisotropic
multichannel single-impurity model also yields a
QCP,\cite{Schlottmann01} while disorder plays an important role in
models based on a single-impurity mechanism,
\cite{Bernal95,Miranda96} in models incorporating interactions
between magnetic ions,\cite{CastroNeto98} and in theories describing
spin fluctuations near a QCP.\cite{Rosch99} The scenario attracting
the most attention at present associates nFL behavior with a nearby
magnetic quantum critical point.\cite{Millis93} A QCP can often be
achieved by doping a pure system chemically, as in the archetype HF
compound UPt$_{3}$ doped with Pd (Ref. \onlinecite{Graf00}),
La-doped CeRu$_{2}$Si$_{2}$ (Ref. \onlinecite{Kambe96}), Au-doped
CeCu$_{6}$, (Ref. \onlinecite{Stockert98}) or Si-doped CeCoGe$_{3}$
(Ref. \onlinecite{Krishnamurtht00}). This introduces additional
disorder and makes the situation even more challenging for theory
because of the need to build a unified picture from the
aforementioned competing mechanisms. Using magnetic field -- where
possible -- as a tuning parameter avoids at least some of the
complications associated with doping. Using the field-tuning
approach has the added advantage of being continuously tunable.
While a QCP, by definition, produces a $T=0$ phase transition, the
finite-temperature properties of the system are also strongly
affected.\cite{Sondhi97} These properties can be examined with a
scaling analysis\cite{Tsvelik93} in order to investigate the nature
of the QCP.

CeCoIn${5}$ is one such system where a QCP can be induced by
magnetic field. Heat capacity\cite{Petrovic01} and de Haas-van
Alphen\cite{Settai01,Hall01} measurements reveal that CeCoIn$_{5}$
is a heavy-electron system. This compound is an ambient-pressure
superconductor with the highest $T_{c}$ (2.3 K) among the Ce-based
HF materials known to-date.\cite{Petrovic01} The unusual magnetic
and thermodynamic properties, both in the normal and superconducting
state, are attracting great interest in this compound. The specific
heat $C$,\cite{Petrovic01,Movshovich01} thermal conductivity
$\kappa$,\cite{Movshovich01} and spin-lattice relaxation time
$T_{1}$ (Ref. \onlinecite{Kohori01}) all display power-law
temperature dependencies below $T_{c}$, while angle-dependent
thermal conductivity\cite{Izawa01} and specific-heat data
\cite{Aoki04} show four-fold modulation. These results indicate that
CeCoIn${_5}$ is quite likely an unconventional line-node
superconductor. In the normal state $C/T$ varies with temperature as
$-$ln $T$ (Ref. \onlinecite{Petrovic01,Nakatsuji02,Bianchi03}),
$1/T_{1}T$ is proportional to $T^{-3/4}$ (Ref.
\onlinecite{Kohori01,Kawasaki03}), $\chi$ varies as $T^{-0.42}$ for
$B$ $\|$ c ,\cite{Kim01}  and the ab-plane resistivity varies
linearly with temperature.\cite{Sidorov02,Nakatsuji02} These nFL
properties have been attributed to the presence of a QCP in the
magnetic phase diagram. No long-range AFM order has been detected in
CeCoIn$_{5}$, although AFM correlations have been
observed\cite{Kohori01}, and these correlations may play a crucial
role in producing the nFL behavior.\cite{Rosch00} Magnetic-field and
temperature-dependent specific-heat\cite{Bianchi03} and ab-plane
transport\cite{Paglione03} measurements suggest that the magnetic
QCP is located close to the upper superconducting critical field,
$B_{c2}$. The origin of the QCP and the nature of the quantum
fluctuations in CeCoIn$_{5}$ are not yet established.

Although CeCoIn$_{5}$ provides a unique opportunity to study nFL
behavior without complications caused by alloying, a careful
separation of co-existing effects is still necessary when analyzing
measured properties. For instance, a systematic study at zero-field
has revealed competing energy scales between single-ion Kondo and
intersite coupling effects.\cite{Nakatsuji02} On the other hand, our
recent in-plane transport study\cite{Hundley04} has shown that the
Hall effect in the Ce-115 materials is strongly influenced by the
conventional electronic-structure that these materials share with
their non-magnetic La-analogs. This suggests that when analyzing
CeCoIn$_{5}$ MR data it is important to account for conventional MR
effects (as determined from LaCoIn$_{5}$ MR data) before associating
any unusual effects with Kondo or QCP physics.

In this paper we present the results of CeCoIn$_{5}$ c-axis
magnetoresistance measurements carried out with the aim of
clarifying the origin of nFL behavior. The measurements were made in
field strengths up to 9 T and at temperatures from 400 K down to 40
mK. In zero field the resistivity along the c-axis, $\rho_{c}$,
varies linearly with $T$ from 40 mK to 8 K and extrapolates
essentially to zero at 0 K. The linear temperature dependence of
both $\rho_{ab}$ and $\rho_{c}$ is consistent with an interplay of
strongly anisotropic scattering due to anisotropic 3D spin
fluctuations and isotropic impurity scattering. Applying a magnetic
field along the c-axis produces a T${^2}$ resistivity, indicating
that FL behavior has been restored. Careful analysis of the
$T^{2}$-coefficient field dependence suggests that the QCP is
located below the upper c-axis critical field, \emph{within} the
superconducting phase. The MR field-dependence below 8 K shows
subtle but important deviations from canonical heavy-fermion
behavior that may be associated with magnetic QCP fluctuations. In
this temperature range the longitudinal magnetoresistance (LMR)
scales with $B$ and $T$ as $B{^2}/T$ due, mainly, to a negative
extrapolated residual resistivity that increases quadratically with
field; at higher temperatures this scaling breaks down, possibly due
to a variation in quenching of Kondo scattering by field for
different charge-carrier bands. For T greater than the coherence
temperature ($\sim$ 45 K) the LMR again shows single-impurity Kondo
behavior; the MR data indicate that the single-ion Kondo scale
$T^{*}$ is roughly 2 K.

\section{\label{Exp}Experiment}
Single crystals of CeCoIn$_{5}$ and LaCoIn$_{5}$ were grown from
an excess In flux, as described in Ref.~[\onlinecite{Petrovic01}].
Excess indium was eliminated by etching the samples in 3:1
HCl:H$_{2}$O solution. CeCoIn$_{5}$ specimens were polished into
rectangular shape, while LaCoIn$_{5}$ samples were left in their
as-grown plate-like shape. All specimens were pre-screened to
ensure that there was no sign of an In superconducting transition
at 3.2 K. Electrical contacts in a standard linear four-probe
configuration were prepared with silver epoxy while silver paste
was used when employing a van der Pauw configuration.

Two CeCoIn$_{5}$ specimens were used in performing anisotropic
$\rho_{xx}(B,T)$ and $\rho_{xy}(B,T)$ (Hall resistivity)
measurements, hereafter denoted as samples I and II. The in-plane
and out-of-plane resistivities of CeCoIn$_{5}$ were determined on
crystallographically oriented sample I via the anisotropic van der
Pauw method.\cite{Pauw58,Price73} Sample I had a thickness of 0.2 mm
and lengths of 0.5 and 0.8 mm along the c-axis and a-axis,
respectively. The measurements in magnetic field were carried out on
sample II (0.1$\times$0.2$\times$0.6 mm$^3$), with the longest
dimension along the c-axis. LaCoIn$_{5}$ samples had thickness
varying from 0.03 to 0.06 mm (along the c-axis) and dimensions of
0.5$\times$1 mm$^{2}$ in the ab-plane. The $\rho_{ab}$ vs. $T$
curves for LaCoIn$_{5}$ were normalized to the average value of the
room-temperature resistivity as determined from anisotropic van der
Pauw measurements.

The temperature and field variation of resistivity from 1.8 K to 400
K and in fields up to 9 T were studied using a Quantum Design PPMS
cryostat while measurements from 40 mK to 2 K were carried out in a
$^{3}$He/$^{4}$He dilution refrigerator. In both cases resistance
measurements were made with an LR-700 ac resistance bridge. The
magnetic field was applied parallel to the current flowing through
the sample. The advantage of using this longitudinal configuration
is that it minimizes or eliminates the influence of ``classical''
magnetoresistance effects arising from the Lorentz force. Magnitudes
of the magnetoresistance reported here are defined in the usual way
as $\Delta\rho/\rho_{o} =[\rho(B)-\rho(B=0)]/\rho(B=0)$.

\section{\label{Res}Results}
\begin{figure}
\includegraphics[width=0.47\textwidth, trim= 5 5 5 5]
{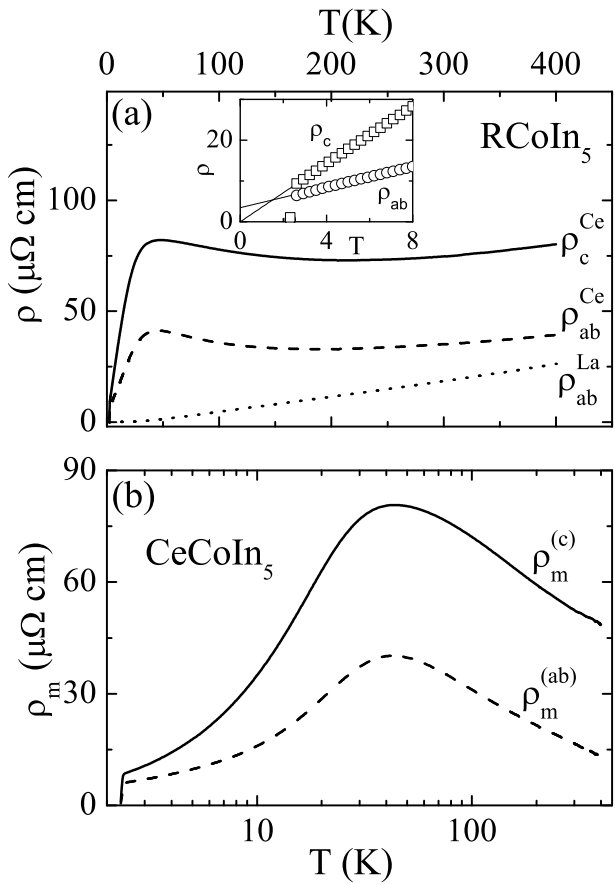}\caption{\label{Resistivity}(a) Resistivity of
CeCoIn$_{5}$ ($\rho_{ab}$ and $\rho_{c}$) and LaCoIn$_{5}$
($\rho_{ab}$) plotted as a function of temperature. The inset shows
the out-of-plane and in-plane resistivities of CeCoIn$_{5}$ for T
$\leq$ 8 K along with linear extrapolations down to zero
temperature. (b) The ab-plane ($\rho_{m}^{ab}$) and c-axis
($\rho_{m}^{c}$) magnetic resistivities of CeCoIn$_{5}$ plotted as a
function of temperature.}
\end{figure}

The out-of-plane ($\rho_{c}$) and in-plane ($\rho_{ab}$)
resistivities of CeCoIn$_{5}$, measured simultaneously on a single
crystal via the anisotropic van der Pauw technique,\cite{Price73}
together with the in-plane resistivity of the non-magnetic analog,
LaCoIn$_{5}$, are depicted in Fig.~\ref{Resistivity} (a). The
in-plane resistivity of LaCoIn$_{5}$ decreases almost linearly
with decreasing temperature and saturates below 10 K to a
sample-dependent value of roughly 0.05 $\mu\Omega$ cm. The
residual resistivity ratio ($\sim$ 350) indicates that the
crystals grown via the flux-growth technique are of high-quality.
We were unable to measure $\rho_{c}$ for LaCoIn$_{5}$ because the
crystals grow as extremely thin plates. An estimate of the
LaCoIn$_{5}$ c-axis resistivity can be determined from
LaRhIn$_{5}$ transport data. For LaRhIn$_{5}$ it was found
previously that the anisotropy ratio $\rho_{c}/\rho_{ab}\sim1.2$
is nearly $T$-independent, suggesting that the inherent
nonmagnetic electronic anisotropy is relatively small for the
RMIn$_{5}$ (R=Ce,La; M=Co,Ir,Rh) structure.\cite{Christianson02}
In the following we assume that $\rho_{c}$ can be quite reasonably
approximate as $1.2\rho_{ab}$ for LaCoIn$_{5}$ as well.

The temperature-dependence of the CeCoIn$_{5}$ zero-field
resistivity is much more complex than that of its nonmagnetic
analog. At room temperature $\rho_{c}$ is 2.1 times larger than
$\rho_{ab}$, indicating that magnetic scattering in CeCoIn$_{5}$ is
modestly anisotropic. Between 400 K and 45 K $\rho_{c}$ and
$\rho_{ab}$ are weakly $T$-dependent, exhibiting a very gradual
minimum centered at roughly 200 K. Below $\sim$45 K the resistivity
in both directions start to decrease rapidly with decreasing $T$.
This behavior is typical for Kondo lattice systems and indicates the
development of Kondo coherence effects.\cite{Grewe91} Below $\sim$10
K $\rho_{c}$ and $\rho_{ab}$ vary linearly with temperature.
Although $\rho_{ab}$ extrapolates to a finite value at zero
temperature (3.8 $\mu\Omega$ cm for the sample I, presented in
Fig.~\ref{Resistivity}), the out-of-plane resistivity of
CeCoIn$_{5}$ extrapolates nearly to zero (see the inset to
Fig.~\ref{Resistivity} (a)).

If we assume that Matthiessen's rule is valid, the magnetic parts of
CeCoIn$_{5}$'s zero-field resistivities in both crystallographic
directions, $\rho_{m}^{(c)}$ and $\rho_{m}^{(ab)}$, can be obtained
by subtracting $\rho_{La}$ from $\rho_{Ce}$. The in-plane and c-axis
magnetic resistivities of CeCoIn$_{5}$ calculated in this manner are
shown in Fig.~\ref{Resistivity} (b). At high temperatures both
$\rho_{m}^{(c)}$ and $\rho_{m}^{(ab)}$ vary as $-$ln$(T)$,
consistent with single-impurity Kondo scattering.\cite{Edelstein03}
Throughout the whole temperature range displayed in the figure
$\rho_{m}^{(c)}$ is higher than $\rho_{m}^{(ab)}$, and the magnetic
anisotropy ratio $r_{m}\equiv\rho_{m}^{(c)}/\rho_{m}^{(ab)}$ drops
with decreasing temperature. The anisotropy ratio is $\sim 3.2$ at
295 K, decreases with decreasing $T$, and reaches a local minimum
($r_{m}=2$) at the temperature where the resistance shows a
coherence maximum (42 K). At still lower temperatures $r_{m}$ drops
gradually, reaching a value of $r_{m}=1.4$ at 2.5 K. Apart from the
$T$-dependent anisotropy $\rho_{m}$ is qualitatively the same in
both directions, with the coherence peak at 42 K.

The differences between the resistivities of LaCoIn$_{5}$ and
CeCoIn${_5}$ are even more evident when measurements are made in a
magnetic field. The T-dependent magnetoresistance of LaCoIn$_{5}$
varies with temperature in a manner typical for a conventional
metal. In the whole $T$ range studied, 2 K -- 300 K, the ab-plane
LMR is positive. Its magnitude at 9 T increases smoothly with
decreasing $T$, changing from a room-temperature value of
$\Delta\rho/\rho_{o} = +0.24\%$ to a 10 K value of
$\Delta\rho/\rho_{o} = +33\%$. Below 10 K the LMR decreases slightly
in magnitude, attaining a 9 T value of $+29\%$ at 2 K. The
transverse magnetoresistance (TMR) in the ab-plane ($B\parallel$ c)
shows the same temperature behavior as the LMR and is roughly ten
times bigger at all temperatures. A more detailed look at the
LaCoIn$_{5}$ MR temperature and field dependence will be presented
in section \ref{La}.

\begin{figure}
\includegraphics[width=0.47\textwidth, trim= 5 5 5 5]
{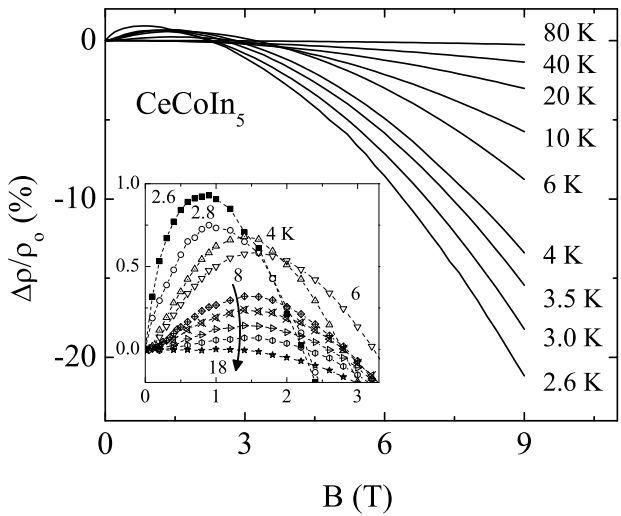}\caption{\label{CeLMR} Longitudinal magnetoresistance of
CeCoIn$_{5}$ plotted as a function of applied field. The low-field
region is magnified in the inset.}
\end{figure}

The field-dependent longitudinal magnetoresistance of CeCoIn$_{5}$
measured at various temperatures with B $||$ c is shown in
Fig.~\ref{CeLMR}; the data differ markedly from that of
LaCoIn$_{5}$. The LMR is positive at 350 K, varies quadratically
with field, and has a 9 T value of $+0.12\%$. The LMR decreases with
decreasing temperature, and it becomes negative at 110 K. It stays
negative in the high field region ($B>3$ T) down to the lowest
temperature in the normal state, with a magnitude that grows with
decreasing $T$; just above $T_{c}$ the MR achieves a 9 T value of
$-21\%$. A positive feature is also evident in the LMR data below 16
K for H $<$ 3 T, as highlighted in the inset to Fig.~\ref{CeLMR}.
Although small compared to the high-field negative
magnetoresistance, the positive feature grows in size with
decreasing temperature, reaching a maximum of $1\%$ just above
T$_{c}$.

\begin{figure}
\includegraphics[width=0.47\textwidth, trim= 5 -5 5 5]
{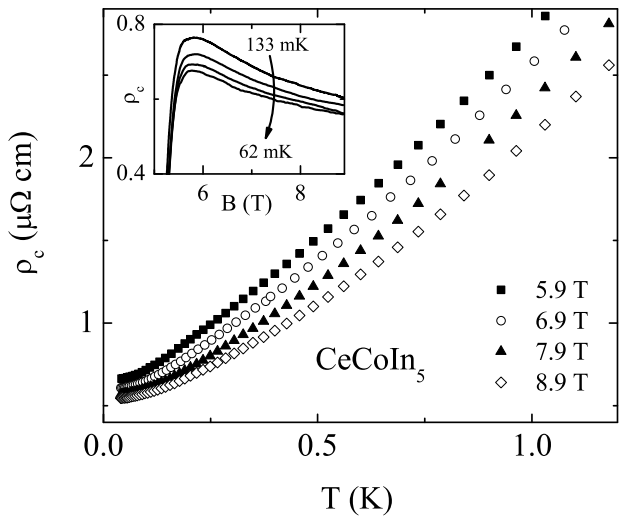}\caption{\label{RFL} Low-temperature c-axis resistivity
of CeCoIn$_{5}$ measured at various magnetic fields as a function of
temperature. The field was applied parallel to the current
direction. The inset shows the c-axis resistivity plotted vs. field
at (from top to bottom): 133, 106, 82 and 62 mK.}
\end{figure}

The most dramatic change in the  resistivity of CeCoIn$_{5}$ is
revealed when  superconductivity is suppressed by applying a
magnetic field. In zero field $\rho_{c}$ varies linearly with
temperature from 8 K down to T$_{c}$, indicative of nFL behavior.
Extending the normal state down to lower $T$ by applying a magnetic
field along the c-axis alters this temperature dependence in an
important way. This is shown in Fig.~\ref{RFL} where $\rho_{c}(T)$
is plotted for fields greater than $B_{c2}$ = 5 T. The curvature in
$\rho_{c}(T)$ visible in the low-$T$ region in Fig.~\ref{RFL}
clearly indicates a departure from a linear $T$-dependence to one of
the form $\rho_{c}\sim T^{n}$, with $n>1$. At 5.9 T, $\rho_{c}(T)$
is proportional to $T^{2}$ below $\sim130$ mK, indicating that a
field-induced FL state has been achieved. The FL regime extends to
0.2 K in a 8.9 T field (see Fig. \ref{T2}). The LMR in this
temperature region is negative in the normal state, as shown in the
inset to Fig.~\ref{RFL}. This differs from the positive MR seen in
low-$T$ ab-plane transport measurements.\cite{Paglione03}

\section{\label{Disc}Discussion}

In analyzing the MR data of CeCoIn$_{5}$ it is very important to
separate field effects associated with many-body or magnetic
interactions from conventional effects intrinsic to the complex
RCoIn$_{5}$ electronic structure. As such, we will start by
analyzing the LaCoIn$_{5}$ magnetoresistance in section \ref{La}. By
separating the MR of CeCoIn$_{5}$ into conventional and magnetic
components we can reveal the presence of a single-impurity Kondo
scale $T^{*}$ in the data; this is discussed in section
\ref{Single}. The zero field resistivity at low temperatures along
the c-axis is analyzed in section \ref{nFL} and compared with
$\rho_{ab}(T)$. The magnetoresistance in the coherence regime is a
subject of section \ref{MnFL}. Lastly, section \ref{FLregime} is
devoted to the restoration of FL behavior by applying a magnetic
field.

\subsection{\label{La}The magnetoresistance of LaCoIn$_{5}$}

Above 20 K the magnetoresistance intrinsic to the RCoIn$_{5}$
electronic structure, given by the MR of LaCoIn$_{5}$, is a
significant part of the total CeCoIn$_{5}$ MR. At 50 K, for example,
the MR of LaCoIn$_{5}$ accounts for roughly 20$\%$ of the total MR
exhibited by CeCoIn$_{5}$ in 9 T. Thus, before analyzing
CeCoIn$_{5}$'s MR, we focus in this section on the magnetoresistance
of its nonmagnetic analog.

\begin{figure}
\includegraphics[width=0.47\textwidth, trim= 5 -5 5 5]
{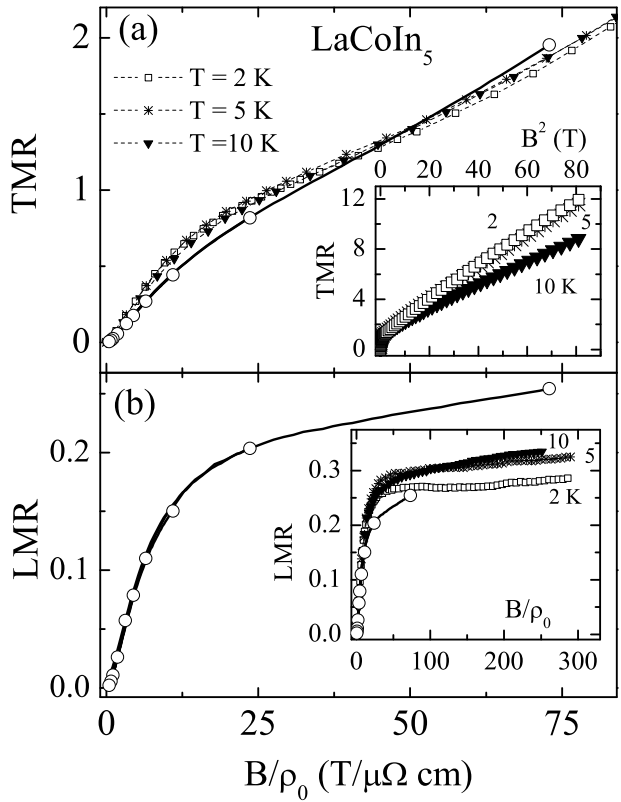}\caption{\label{KohlerPlot} (a) Transverse ($\vec B
\parallel$ c) and (b) longitudinal ($\vec B \parallel$ ab-plane and
current) ab-plane magnetoresistance of LaCoIn$_{5}$ as a function
of $B/\rho(B=0)$. The open circles in both panels show the 9 T
 magnetoresistance for (from left to right) $T=$ 300, 200, 150, 100,
 70, 60, 50, 40, 30 and 20 K, respectively. The solid line in both
 panels correspond to field-swept (0 to 9 T) MR data taken at the
 temperatures listed above; these field-sweeps are
 indistinguishable from one another, indicating that the MR of
 LaCoIn$_{5}$ obeys Kohler's rule. MR data at 2, 5, and 10 K are
 also shown in (a) and the inset to (b) as a function of $B/\rho(B=0)$.
 The inset to (a) shows the transverse MR at 2, 5, and 10 K as
 a function of B$^{2}$.}
\end{figure}

The transverse and longitudinal ab-plane MR of LaCoIn$_{5}$ measured
at temperatures from 300 K down to 20 K and in fields to 9 tesla are
plotted as a function of B divided by the zero-field resistivity
($\rho_{0}$) in Fig. \ref{KohlerPlot}. The data clearly collapse
onto a common curve, indicating that Kohler's rule,\cite{Ziman60}
\begin{equation}
\frac{\Delta\rho}{\rho_{0}}=F\left(\frac{B}{\rho_{0}}\right),
\label{Kohler}
\end{equation}
(where $F(x)$ is an unspecified function that depends on details of
electronic structure) is fulfilled in LaCoIn$_{5}$ over a wide range
of $T$ encompassing an almost two order-of-magnitude variation in
$\rho_{0}$. Changes in temperature evidently alter the magnitude of
the relaxation time, $\tau$, by the same factor for all electron
wave vectors, $\vec k$, without altering the form of $\tau(\vec
k)$.\cite{Chambers56} The literature on magnetoresistance in metals
focuses far more attention on applying Kohler's rule to the
transverse configuration, and all but ignores the longitudinal
configuration. It is worth noting, however, that the scaling
argument describing the way in which the trajectory of the charge
carriers is altered when the field $B$ and the scattering rate
1/$\tau$ are simultaneously increased by the same factor can be
applied to \textit{any} measured resistivity. As such,
Eq.~\ref{Kohler} can, in principle, be applied to \textit{any}
component of the resistivity tensor.\cite{Pippard89} Deviations from
Kohler's rule are evident in the LaCoIn$_{5}$ MR data below 20 K.
This is the same $T$ region where the $\rho(T)$ curve becomes
saturated, i.e. where residual impurity scattering begins to
dominate electron-phonon scattering. A change in the dominant
scattering mechanism leads, presumably, to an alteration in
$\tau(\vec k)$ below 20 K, resulting in modest deviations from
Kohler's rule.\cite{Chambers56}

As shown in the insets to Fig. \ref{KohlerPlot} the low
temperatures/high-field ab-plane MR of LaCoIn$_{5}$ becomes strongly
anisotropic, and the behavior in the high-field limit reflects the
underlying Fermi surface topology intrinsic to the ``115''
structure. The LMR (H $\perp$ c) at 2, 5 and 10 K saturates in high
fields while the TMR (H $\parallel$ c) increases approximately as
$B^{2}$ without any sign of saturation when measured in fields up to
9 tesla. In a compensated metal (such as LaCoIn$_{5}$) where the
area of hole and electron Fermi surfaces are equal, the TMR is
expected to vary quadratically with B in the high-field limit when
all orbits in planes normal to the applied field direction are
closed.\cite{Hurd73} The electron and hole Fermi surfaces in La-115
and Ce-115 materials are very complex,\cite{Shisido02,Hall01,Haga01}
and both de Haas-van Alphen (dHvA) measurements and band-structure
calculations indicate that the complex FS topology of LaRhIn$_{5}$
is dominated by corrugated electron-like cylindrical orbits that run
along the c-axis.\cite{Shisido02} In such a situation the orbits in
the ab-plane ($\vec B\parallel$ c-axis) are indeed closed, offering
a simple explanation\cite{closedorbits} for both the
$B^{2}$-dependence of the ab-plane TMR and the linear $B$-dependence
of the Hall voltage reported recently.\cite{Hundley04} For a
magnetic field applied perpendicular to the axis of a corrugated
cylinder, the LMR is expected to saturate in the high-field
limit.\cite{openorbits} This tendency is observed in the
LaCoIn$_{5}$ MR when a field is applied in the ab-plane (see the
inset to Fig.~\ref{KohlerPlot}(b)). Hence, the directional
dependence of the low-$T$ MR in LaCoIn$_{5}$ is consistent with the
cylindrical Fermi surface topology observed in dHvA measurements.

\subsection{\label{Single}Single impurity regime of CeCoIn$_{5}$.}
By utilizing our knowledge of the MR of LaCoIn$_{5}$ it is possible
to examine the CeCoIn$_{5}$ magnetoresistance components that stem
from Kondo or other magnetic interactions. Again, assuming that
Matthiessen's rule is valid at finite field, we can simplify the MR
problem by decomposing the total $B$-dependent resistivity of
CeCoIn$_{5}$, $\rho_\text{tot}(B)$, into two independent parts:
$\rho_\text{tot}(B)=\rho_\text{mag}(B)+\rho_\text{La}(B)$;
$\rho_\text{mag}(B)$ here is the magnetic-scattering contribution to
the overall resistivity. We assume that the contribution of all
other mechanisms can be approximated by the field-dependent
resistivity of LaCoIn$_{5}$, $\rho_\text{La}(B)$. The similarity
between the electronic structures of CeRhIn$_{5}$ and LaRhIn$_{5}$
as observed in dHvA measurements\cite{Shisido02,Hall01} corroborates
this supposition. Next we define the magnetic part of the
magnetoresistance, MR$_{\text{mag}}$, as

\begin{equation}
\text{MR}_{\text{mag}}=\frac{\Delta\rho_{mag}(B)}{\rho_{mag}(0)}=\frac
{\rho_{mag}(B)-\rho_{mag}(0)}{\rho_{mag}(0)}. \label{mMR}
\end{equation}

In determining the magnetic longitudinal magnetoresistance of
CeCoIn${_5}$ (LMR$_{mag}$) we are forced to infer the c-axis B and
T-dependent resistivity of LaCoIn$_{5}$ from ab-plane data because
the La-analog sample thickness precludes making c-axis transport
measurements. As discussed in the previous section, Fermi surface
anisotropy only influences the magnetoresistance of LaCoIn$_{5}$
in the low-$T$ region. As such it is reasonable to assume that the
c-axis and ab-plane magnetoresistance of LaCoIn$_{5}$ will be
similar above $\sim$ 20 K.

\begin{figure}
\includegraphics[width=0.47\textwidth, trim= 5 5 5 5]
{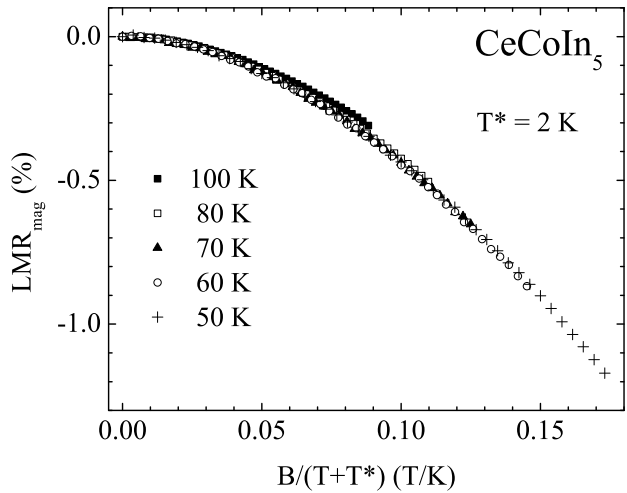}\caption{\label{SingleImpurity} f-electron contribution
to the CeCoIn$_{5}$ longitudinal MR as a function of $B/(T+T^{*})$,
with $T^{*} = 2$ K.}
\end{figure}

By following this recipe we find that the magnetic contribution to
the longitudinal MR of CeCoIn$_{5}$ is negative below 200 K,
varies quadratically with field, and grows in magnitude with
decreasing temperature; for example, in a field of 9 tesla, the
magnetic contribution to the longitudinal MR is $-0.3\%$ and
$-1.2\%$ at 100 and 50 K, respectively. A negative MR that grows
with decreasing temperature and increasing field is consistent
with single-ion Kondo behavior.\cite{Schlottmann89} Thus it is
natural to carry out a scaling analysis of the LMR$_{mag}$ data as
suggested by the Bethe-ansatz solution of the Coqblin-Schrieffer
model.\cite{Coqblin69,Schlottmann89} In this single-impurity Kondo
model the relative magnetoresistance depends on $B$ and $T$ only
through the ratio $B/(T+T^{*})$,
\begin{equation}
\frac{\Delta\rho(B,T)}{\rho(0,T)}=f\left(\frac{B}{T+T^{*}}\right),
\label{scalling}
\end{equation}
where $T^{*}$ plays the role of the single-ion Kondo temperature. In
Fig.~\ref{SingleImpurity} we show that all LMR$_{mag}(B)$ curves
between 50 K and 100 K can be superimposed onto a single unique
curve when the data are scaled according to Eq.~(\ref{scalling}).
Scaling works best for $T^{*}=(2 \pm 2)$ K. Although this value is a
relatively small number when compared with the temperature range of
interest, the quality of the scaling overlap begins to deteriorate
when $T^{*}$ is changed to values greater than 4 K. The
temperature-range over which the MR data scale coincides roughly
with the region of $-\text{log }T$ behavior exhibited by $\rho_{m}$
(see Fig.~\ref{Resistivity}). The presence of a small
single-impurity Kondo energy scale of roughly 1 to 2 K was reported
in a systematic study of the zero-field resistivity, magnetic
susceptibility, and specific heat of
Ce$_{\text{1-x}}$La$_{\text{x}}$CoIn$_{5}$ (Ref.
\onlinecite{Nakatsuji02}). The Kondo energy-scale was found to be
essentially constant from the dilute limit ($x\rightarrow1$) to the
Kondo lattice limit ($x\rightarrow0$). Hence, by properly accounting
for conventional non-magnetic contributions to the MR, we are able
to discern the single-impurity energy scale in the overall
magnetotransport properties of CeCoIn${_5}$.

\subsection{\label{nFL}Transport at zero field in non-Fermi-liquid regime }

In this section we examine the influence of AFM-fluctuations,
dimensionality, and disorder in the low temperatures zero-field
transport of CeCoIn$_{5}$ where in-plane\cite{Paglione03,Bianchi03}
and out-of-plane resistivity data clearly show evidence of nFL
behavior. We fit the low-$T$ data to the form
\begin{equation}
\rho(T)=\rho_{0}+AT^{n} \label{quasiLinear}
\end{equation}
by plotting the data as $(\rho-\rho_{0})/T^{n}$ vs. $T$ and
adjusting $n$ and $\rho_{0}$ to produce a horizontal line. As
pointed out previously,\cite{Sidorov02} this approach produces
fitting parameters that are far less sensitive to the temperature
range under consideration than when directly fitting the data to Eq.
\ref{quasiLinear}.

\begin{figure}
\includegraphics[width=0.47\textwidth, trim= 5 5 5 5]
{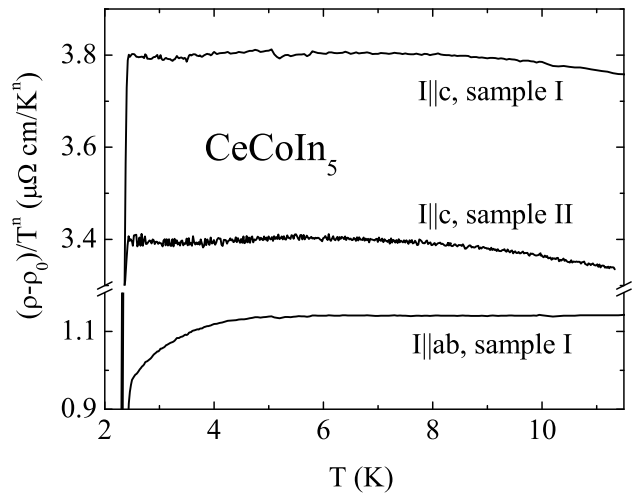}\caption{\label{RhoN} Zero field resistivity of
CeCoIn$_{5}$ plotted as $(\rho-\rho_{0})/T^{n}$ vs. $T$. The upper
and lower curves show $\rho_{ab}$ and $\rho_{c}$ of sample I as
measured via the anisotropic van der Pauw method, while the middle
curve depicts $\rho_{c}$ for sample II. The best fit for the
ab-plane data (corresponding to a horizontal line) occurs for
$\rho_{0}=3.8$ $\mu\Omega$ cm, while fits to the c-axis data leads
to slightly negative $\rho_{0}$ values: $\rho_{0}$ = $-0.30$, and
$-0.39$ $\mu\Omega$ cm for sample I, and II, respectively}.
\end{figure}

Results for CeCoIn$_{5}$ samples I and II are plotted in
Fig.~\ref{RhoN}. Simultaneous measurements of $\rho_{c}(T)$ and
$\rho_{ab}(T)$ on sample I produce fitting exponents that are
essentially equivalent: $n_{ab} = 1.03 \pm 0.02$ and $n_{c} = 0.97
\pm 0.02$. The ab-plane resistivity begins to deviate from this
linear T-dependence above 12 K, while the c-axis data show a similar
deviation starting at 8 K. The ab-plane data also deviate from the
horizontal trend for $T < 4$ K, well above $T_{c}$. A similar
deviation, although at a slightly lower temperature ($\sim3.4$ K),
was observed previously\cite{Sidorov02} and attributed to the
opening of a pseudogap. The existence of such a gap is still subject
to debate\cite{Sidorov02} since specific heat and magnetic
susceptibility measurements have yet to produce confirming evidence
that it exists. As indicated in Fig.~\ref{RhoN}, c-axis measurements
on sample II give practically the same result as for sample I, with
a power-law exponent of $n_{c} = 1.00 \pm 0.02$. The data therefore
reveal that the out-of-plane resistivity in CeCoIn$_{5}$ changes
linearly with temperature between $T_{c}$ and roughly 8 K and show
no evidence for a pseudogap.

The spin fluctuation (SF) theories of non-Fermi-liquid behavior
predict $n=1$ and $n=1.5$ for two-dimensional and three-dimensional
quantum-critical (QC) systems,
respectively.\cite{Hertz76,Millis93,Moriya95,Lonzarich97,Sachdev99}
Recent In-NQR and Co-NMR measurements indicate that the AFM spin
fluctuations in CeCoIn$_{5}$ are 3D with anisotropy such that the
magnetic correlation length along the c-axis is shorter than that
within the tetragonal plane.\cite{Kawasaki03} If AFM spin
fluctuations associated with a QCP are responsible for the T-linear
resistivity exhibited by CeCoIn$_{5}$, then correlation-length
anisotropy provides a simple explanation of why the temperature
region of linear $\rho(T)$ dependence for in-plane transport is
larger than for transport along the c-axis.

The discrepancy between the observed temperature exponent ($n=1$)
and that expected for 3D system ($n=1.5$) can be clarified by taking
into account the role of disorder in a 3D system. When the dual
effects of isotropic impurity scattering and anisotropic
spin-fluctuation scattering on $\rho(T)$ are calculated for a 3D
system, $\rho\propto T^{1.5}$ behavior is only realized at very low
temperatures on the order of $10^{-3}\Gamma$, where $\Gamma$ is a
characteristic SF energy scale.\cite{Rosch99,Rosch00,Rosch01} For HF
systems $\Gamma$ is comparable to the coherence temperature,
$T_{coh}$.\cite{Rosch00} CeCoIn$_{5}$ resistivity data indicate that
$T_{coh}\approx45$ K, so that the aforementioned very-low-$T$ region
(T $<$ 45 mK) is not accessible due to the 2.3 K superconducting
transition. In the experimentally accessible
intermediate-temperature region transport exponents near 1.0 are
expected for a clean system.\cite{Rosch99} Following Rosch, the
inverse of the residual resistivity ratio can serve as an estimate
of the degree of disorder, $x \approx
\rho_{ab}(T\rightarrow0)/\rho_{ab}(300 K)$.\cite{Rosch00} According
to this criterion sample I with $x\approx 0.1$ is relatively clean
and $n\sim1$ is expected in the temperatures of the order of
$\sim0.1\Gamma$,\cite{Rosch99} and this is what we observed
experimentally.

\subsection{\label{MnFL}The magnetoresistance of CeCoIn$_{5}$
in the coherence regime}

In this section we discuss the LMR of CeCoIn$_{5}$ for temperatures
below the point where the zero-field resistivity exhibits a
coherence peak ($T_{coh} \sim 45$ K). The data plotted in
Fig.~\ref{CeLMR} clearly indicate that the LMR is generally negative
and grows in magnitude with decreasing temperature. For $T <
T_{coh}$ the zero-field resistivity of LaCoIn$_{5}$ is minuscule
compared to the resistivity of CeCoIn$_{5}$ ($\rho_{La}/\rho_{Ce}
\sim0.2\%$), and their ratio is essentially unchanged even in 9 T.
As such, any conventional electronic-structure contribution to the
magnetoresistance of CeCoIn$_{5}$ is negligible, and the MR of
CeCoIn$_{5}$ below 20 K can be fully attributed to the presence of
Ce ions and f-electrons.

\begin{figure}
\includegraphics[width=0.47\textwidth, trim= 5 -5 5 5]
{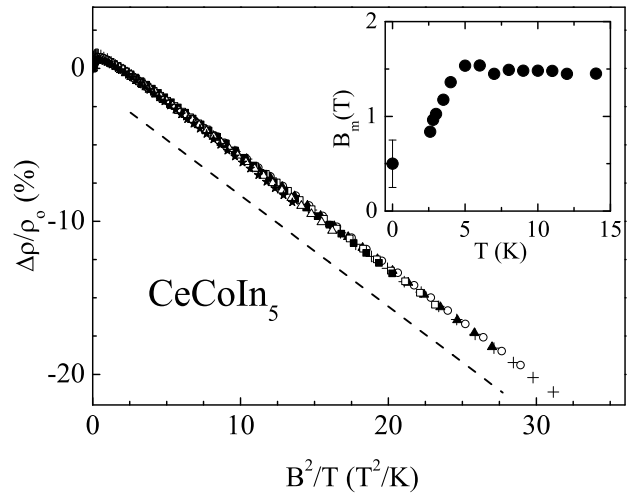}\caption{\label{MRscale} LMR isotherms for CeCoIn$_{5}$
at 2.6, 2.8, 3.0, 3.5, 4, 5 and 6 K plotted as a function of
$B^{2}/T$; the dashed line is a guide to the eye. The inset shows
the field of the LMR maximum $B_{m}$ as a function of temperature.
The $B_{m}$ value at $T = 0$ comes from the zero-$T$ extrapolations
of LMR($T$) data taken at different fields.}
\end{figure}

\begin{figure}
\includegraphics[width=0.47\textwidth, trim= 5 5 5 5]
{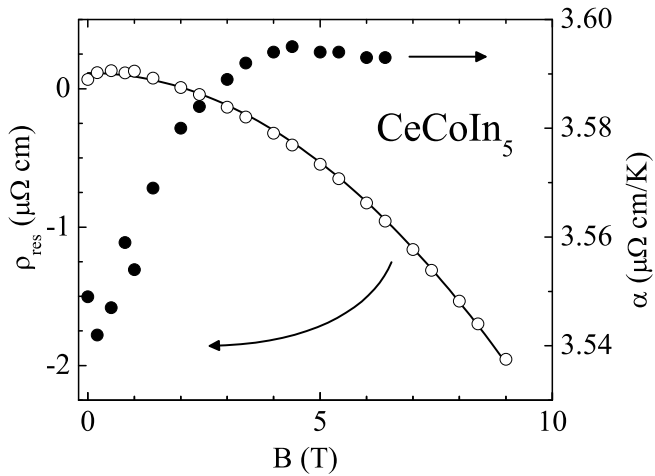}\caption{\label{nFLparams} Fitting parameters from
constant-field CeCoIn${_5}$ data for $1.8K \leq T \leq 8 K$ plotted
as a function of field. The solid line is a quadratic fit to
$\rho_{res}(B)$ with a small $B = 0$ offset.}
\end{figure}

We focus first on the negative contribution to the low-temperature
LMR that dominates the data depicted in Fig. \ref{CeLMR} for $B > 3$
tesla. This negative component varies quadratically with field and
grows rapidly with decreasing temperature. As shown in Fig.
\ref{MRscale} the LMR field sweeps for temperatures ranging from 6 K
down to $T_{c}$ can be superimposed onto a common line when the data
are replotted as a function of the scaling parameter $B{^2}/T$. The
c-axis resistivity shows nFL behavior ($\rho_{c} \propto T$) at the
temperatures where the MR data scale in this way. The cause for the
$B{^2}/T$ LMR scaling in the nFL regime becomes clear when we
parameterize the resistivity through the expression
\begin{equation}
\rho(B,T) = \rho{_{res}}(B) + \alpha(B)T. \label{nFLLMRfit}
\end{equation}
The $\rho(B,T)$ data vary linearly with $T$ from 1.5 K up to roughly
8 K, and the field dependence for $\rho_{res}$ and $\alpha$ as
extracted from linear fits to the $\rho(B,T)$ data are shown in Fig.
\ref{nFLparams}. The slope is weakly field-dependent, changing by
only $1.1\%$ when the field is increased from 0 to 9 tesla. In
contrast, the extrapolated residual resistivity term is quite
field-dependent; beginning at $B = 0$ where $\rho_{res}$ is
essentially zero, the residual term becomes increasingly negative as
$B$ is increased, and it varies quadratically with the field
strength. The $\rho(B,T)$ parametrization shown in Fig.
\ref{nFLparams} indicates that the applied field serves to offset
the nFL resistivity downward. The $B^{2}/T$ LMR scaling directly
follows from (1) a field-independent slope in $\rho(T)$, (2) a
negligible residual resistivity at $B = 0$, and (3) $\rho_{res}
\propto -B^{2}$ at higher fields. The increasingly negative
$B$-dependent residual resistivity evident from ~1.5 K to 8 K is an
indication that the resistivity must become a stronger function of
$T$ ($\rho \propto T^{n}$ with $n > 1$) at lower temperatures.
Hence, the negative $\rho_{res}(B)$ term in the nFL state simply
reflects the fact that the magnetic field pushes the system into a
FL state at much lower temperatures. This point will be discussed
further in Sec. \ref{FLregime}.

We turn now to the low-field range where a \textit{positive} LMR is
observed and the aforementioned $B^{2}/T$ scaling no longer holds.
As shown in the inset to Fig. \ref{CeLMR} this small positive LMR
appears beginning at roughly 16 K with a magnitude that increases
with decreasing $T$, reaching a maximum value of $\sim1\%$ near the
onset of superconductivity. A low-field positive magnetoresistance
is a common attribute of Kondo-lattice systems in or close to their
Fermi-liquid ground-state. This behavior is exhibited, for example,
by CeRu$_{2}$Si$_{2}$ (Ref. \onlinecite{Flouquet86}), CeAl$_{3}$
(Ref. \onlinecite{Roesler90,Roesler92}), YbNi$_{2}$B$_{2}$C (Ref.
\onlinecite{Yatskar99}), and CeRhIn$_{5}$ (Ref.
\onlinecite{Christianson02}). In a Kondo-lattice Fermi-liquid the MR
maximum results from the competition between a $T$-independent
residual resistivity contribution that increases in a magnetic
field, and a temperature-dependent term that decreases in a magnetic
field and grows quadratically with
temperature.\cite{Ohkawa90,Chen93} However, for the aforementioned
Kondo-lattice case the location, $B_{m}$, of the MR maximum moves
toward lower fields with increasing $T$, as illustrated in Fig.~3 of
Ref.~\onlinecite{Roesler90}. This occurs because the negative MR
component stemming from charge fluctuations grows with increasing
temperature. In CeCoIn$_{5}$ the opposite trend is observed up to 6
K -- the maxima shift toward \textit{higher} fields with higher $T$,
as can be seen in the inset to Fig.~\ref{CeLMR}. The positions of
these maxima, $B_{m}$, obtained from polynomial fits to the
low-field LMR data, are shown in the inset to Fig.~\ref{MRscale}.
The difference between the low-field MR in a coherent Kondo system
and that of CeCoIn$_{5}$ resides in the fact that the extrapolated
residual resistivity term in CeCoIn$_{5}$ produces a negative MR
while a small positive MR results from the slight increase in the
slope of $\rho(T)$ shown in Fig. \ref{nFLparams}. This positive
component grows relative to the negative residual term with
increasing temperature, resulting in $B_{m}$ moving to higher fields
as the temperature is increased. The field-dependent evolution of
the MR in CeCoIn$_{5}$ is quite different from that of a Kondo
system that does not show nFL behavior in the low-$T$ resistivity.
As such the low-field positive LMR in CeCoIn${_5}$ is consistent
with field quenching of the AFM spin fluctuations responsible for
the nFL behavior.

At roughly 6 K a significant change in the LMR behavior takes place;
$B_{m}$ becomes $T$-independent above 6 K and the data no longer
follow the $B^{2}/T$ scaling relationship. Attempts to find any
simple MR scaling in the range 7 K--20 K were unsuccessful. To
clarify the possible origin of this LMR behavior we carried out Hall
effect measurements on sample II below 20 K with $I\|$c and the Hall
voltage $V_{xy}$ measured in the ab-plane. Two characteristic
features are present in the data: first, $V_{xy}$ varies nonlinearly
with field and it changes sign as well. Second, constant-temperature
$V_{xy}(B)$ curves shift toward lower values with increasing $T$ up
to 6 K, and then they start to move in the opposite direction -- to
higher values -- above 6 K. With regard to the second effect, a
shallow minimum in the Hall coefficient $R_{H}(T)$ centered at
roughly the same temperature has been seen in ab-plane transport
measurements when the field exceeded 0.5 T.\cite{Hundley04} This
temperature dependence is typical for multi-band electronic
structure system in which the weighted contribution from different
bands changes with temperature.\cite{Hurd73} The nonlinear field
dependence of $V_{xy}$ can be attributed to the Kondo interactions
present in the system. When an external field is applied to a HF
system, the Kondo resonance will broaden, split, and ultimately
shifts below the Fermi energy.\cite{Satoh85} When two bands are
present, the response to an applied field can be different for the
carries in these two bands. The band for which the field suppresses
the Kondo effect more efficiently will carry a larger fraction of
the aggregate transport current as the field is increased. It seems
quite reasonable that this mechanism can explain the observed change
in the sign of the constant-temperature Hall voltage with increasing
field, while a subtle interplay between Kondo interactions and
multiple bands with different carrier-mobility $T$-dependencies
could be responsible for the change in the LMR field dependence for
$T \geq 6$ K.

\subsection{\label{FLregime}Restoration of Fermi-liquid behavior
by magnetic field}

\begin{figure}
\includegraphics[width=0.47\textwidth, trim= 5 5 5 5]
{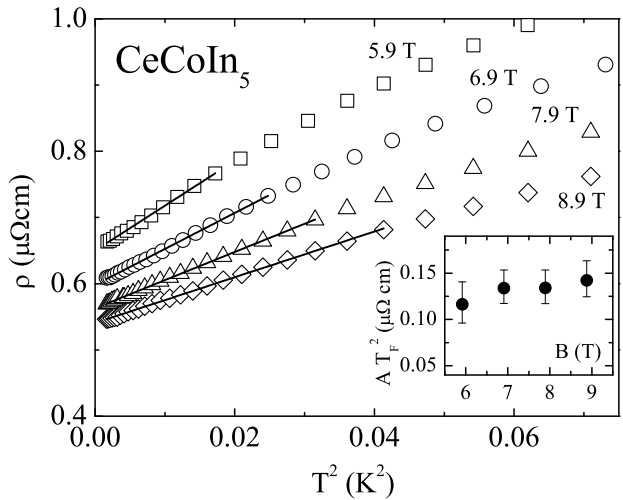}\caption{The c-axis resistivity of CeCoIn$_{5}$ in the
mK range plotted as a function of $T^{2}$ for fields applied
parallel to the c-axis. Solid lines are linear fits from 40 mK up to
$T_{FL}(B)$. The inset shows the product $AT_{FL}^{2}$ (in
$\mu\Omega$ cm) plotted as a function of the field strength in
tesla.\label{T2}}
\end{figure}

This section is devoted to an analysis of the field-induced FL
behavior evident in $\rho_{c}(T)$ data for large magnetic fields.
Applying a magnetic field along the c-axis causes a dramatic change
in the low-temperature $T$-dependence of $\rho_{c}$ in the normal
state. At 5.9 tesla the resistivity data  below 130 mK can be
described by the expression
\begin{equation}
\rho(B,T)=\rho_{0}(B)+A(B)T^{2},
\label{Tsquare}
\end{equation}
characteristic of FL behavior. The temperature range of FL behavior
becomes larger with increasing field strength. The upper limit of
the range where Eq.~\ref{Tsquare} is valid may be roughly identified
as a characteristic temperature, $T_{FL}$, for the onset of
Fermi-liquid behavior. $T_{FL}$ was determined at each field using a
procedure described in Ref.~\onlinecite{Dickey97} that works as
follows: a straight line was fit to the first five $\rho_{c}$ vs.
$T^{2}$ data points beginning at 40 mK and the resulting reduced
$\chi^{2}$ error was calculated. The procedure was repeated by
including successive data points at higher temperature, and $T_{FL}$
was determined as the temperature $T_{\chi}$ where $\chi^{2}$ starts
to grow rapidly. Plots of $(\rho-\rho_{0})/T^{2}$ at different field
strengths were also used to determine $T_{FL}$ as precisely as
possible. In the Landau FL theory the coefficient $A$ in
Eq.~\ref{Tsquare} is inversely proportional to the square of the
characteristic temperature governing the FL behavior. The calculated
product $AT_{FL}^{2}$ is depicted in the inset to Fig.~\ref{T2}; the
error bars reflect the uncertainty in determining the position of
$T_{\chi}$ following the procedure described above. Within these
error bars $AT_{FL}^{2}$ is constant, confirming the consistency of
our analysis.

\begin{figure}
\includegraphics[width=0.47\textwidth, trim= 5 5 5 5]
{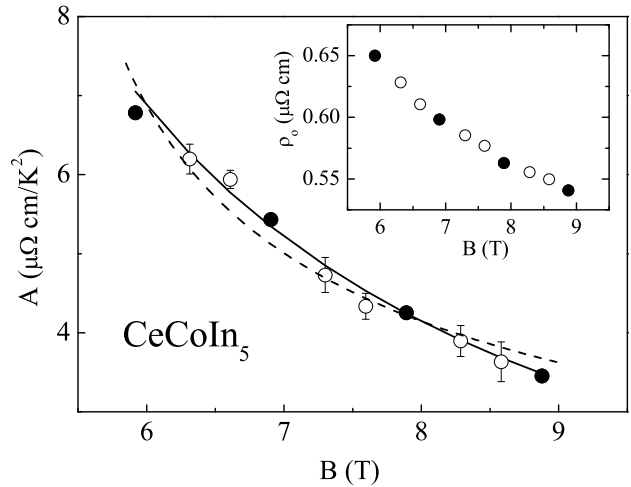}\caption{\label{FL}The T$^{2}$ resistivity coefficient
$A$ plotted as a function of field. The solid and open symbols
correspond to temperature-sweep and field-sweep data, respectively.
The solid line shows the least-squares fit to Eq.~\ref{critical}
when setting $p=1.37$; the fit gives $B_{cr}= 1.5\pm0.2 $ T. The
dashed line shows the fit when setting $B_{cr}=B_{c2}$.
Field-dependent $\rho_{0}$ values are shown in the inset.}
\end{figure}

We now use the field dependence of the coefficient $A$ to determine
the location of the QCP in CeCoIn$_{5}$ in relation to the c-axis
upper critical field $B_{c2} = 4.95$ tesla.\cite{Murphy02,Ikeda01}
Values of $A$ at different fields, taken both from fits to
$\rho_{c}(T)$ data taken at various fields and from $\rho_{c}(B)$
data taken at various temperatures, are shown in Fig.~\ref{FL}. The
data indicate that $A$ is a decreasing function of field, an
entirely expected result given that this coefficient is a measure of
the strength of quasi-particle--quasi-particle interactions, and, as
such, is proportional to the effective mass. We fit the data to a
formula that models the diverging behavior of $A$,

\begin{equation}
A(B) = \frac{{A_0 }}{{\left( {B - B_{cr} } \right)^p }}\;,
\label{critical}
\end{equation}
where $B_{cr}$ is the critical field where A diverges, $p$ is the
critical exponent ($p> 0$), and $A_{0}$ is a constant. The dynamic
range over which $A(B)$ can be measured is limited because the
superconducting ground-state masks the FL transport behavior for
fields less than roughly 6 tesla. For this reason determining
$B_{cr}$ and $p$ unambiguously requires care when analyzing the
data. A plot of $1/A$ vs. $B$ shows upward curvature, indicating
that $p > 1$. An upper limit on $B_{cr}$ can be obtained if we fix
$p$ to 1 and perform a non-linear least-squares fit to the data;
this approach gives $B_{cr} = (3 \pm 0.2) $ tesla, a value that is
clearly less than $B_{c2}$. If we use $p = 1.37$ as determined from
ab-plane transport\cite{Paglione03} and specific-heat
measurements,\cite{Bianchi03} the best-fit to the data (the solid
line in Fig. \ref{FL}) gives $B_{cr} = (1.5 \pm 0.2)$ tesla, putting
the QCP even farther into the superconducting state. If,
alternatively, we force the critical field to coincide with
$B_{c2}$, the resulting best-fit critical exponent ($p = 0.5$)
provides a very poor description of the data. This fit, depicted in
Fig. \ref{FL} as a dashed line, has a $\chi^{2}$ error 10 times
greater than that for the fit that gives $B_{cr} = 1.5$ tesla.
Clearly, the c-axis transport data are inconsistent with $B_{cr}$
being close to $B_{c2}$, but instead places the QCP well inside the
superconducting phase. In a very qualitative sense the negative
values of $\rho_{res}(B)$ obtained for fields greater than roughly 2
tesla (see Fig. \ref{nFLparams}) are also consistent with this
conclusion. Interestingly, a linear fit to a plot of $\rho_{0}/A$
vs. $B$ gives a zero-intercept at $B_{cr}= (1.7\pm0.4$) T, i.e. at
the same field (within the error bars) as $B_{cr}$ determined from
the fit with $p=1.37$. As shown in the inset to Fig.~\ref{RFL} the
magnetoresistance is affected by superconducting fluctuations for
fields below 5.9 T. This precludes enhancing the dynamic range of
the data by determining A at lower fields. Despite the limited field
range used in the analysis, c-axis transport data clearly suggest
that the critical point resides well inside the superconducting
phase.

This conclusion appears to be at odds with ab-plane transport and
specific heat
measurements\cite{Paglione03,Bianchi03,Bauer05,Ronning05} which show
rather clearly that the critical point coincides with B$_{c2}$.
Those measurements indicate that, despite a factor of 2.4 difference
in $B_{c2}$ for $B \perp c$ and $B \parallel c$, $B_{cr}$ tracks
$B_{c2}$ for either field
direction.\cite{Paglione03,Bianchi03,Ronning05} Even more compelling
is the fact that $B_{cr}$ still coincides with $B_{c2}$ when the
critical field is reduced by $50\%$ through Sn doping.\cite{Bauer05}
These results indicate that it is more than just a coincidence that
the critical field occurs at $B_{c2}$. Why, then, do c-axis
magnetotransport data place $B_{cr}$ far below $B_{c2}$? The
complexities intrinsic to the electronic structure of CeCoIn$_{5}$
may be responsible. Band structure calculations and de Haas-van
Alphen measurements indicate the Fermi surface of CeCoIn$_{5}$ is
composed of 3D hole pockets and c-axis oriented 2D electron-like
sheets.\cite{Settai01,Shisido02} While ab-plane transport involves
carriers on both pieces of the Fermi surface, c-axis transport will
be carried predominately by the 3D pockets. Given the large
anisotropy in $B_{cr}$ and $B_{c2}$, it is not unreasonable to
conclude that critical fluctuations are more prevalent on the 2d
sheets. If true, c-axis transport would not be heavily influenced by
fluctuations associated with the QCP, but would instead reflect a
more complicated mix of field-dependent transport effects. These
magnetotransport complications could alter our critical point
analysis sufficiently to mask the true location of $B_{cr}$.

\section{\label{Concl}Conclusions}

The c-axis transport of CeCoIn$_{5}$ is dominated by single-impurity
Kondo scattering at high temperatures while AFM critical
fluctuations associated with a nearby QCP control the transport at
low temperatures. Between 50 and 100 K the longitudinal
magnetoresistance of this Kondo lattice compound is consistent with
a single-impurity Kondo energy scale of roughly 2 K. Below 10 K the
$T$-linear nFL behavior of both $\rho_{ab}$ and $\rho_{c}$ are
consistent with anisotropic 3D AFM spin fluctuations in a relatively
clean system. As in previous ab-plane studies,\cite{Paglione03}
applying a magnetic field along the c-axis restores FL behavior at
low temperatures. In sharp contrast to those ab-plane
magnetotransport measurements, the field dependence of $\rho_{c}$ in
the field-induced FL regime suggests that the QCP in CeCoIn$_{5}$
resides well inside the superconducting phase; this result is at
odds with a number of ab-plane transport and thermodynamic
measurements which place the critical point at $B_{c2}$. The
magnetic fluctuations associated with the QCP influence the
transport properties at least up to 16 K. The influence that these
fluctuations have on the electronic transport are reduced by
increasing the temperature or applying a magnetic field to the
system. For large fields the LMR becomes negative as the system is
pushed away from the QCP. Changes in the LMR field dependence above
6 K suggest that the complex multiband electronic structure strongly
influences the $B$-dependent electronic transport in CeCoIn$_{5}$.

\acknowledgments We thank Z. Fisk and J. Lawrence for useful
discussions. This work was performed under auspices of the U.S.
Department of Energy.

\end{document}